\documentstyle[12pt,aasms4]{article}
\begin{document}
\def\mv{M_V}
\def\mi{M_I}
\def\mj{M_J}
\def\MH{{\rm [M/H]}}
\def\mvol{M_\odot pc^{-3}}
\def\msurf{M_\odot pc^{-2}}
\def\msol{M_\odot}
\def\zsol{Z_\odot}
\def\te{T_{eff}}
\def\simgr{\,\hbox{\hbox{$ > $}\kern -0.8em \lower 1.0ex\hbox{$\sim$}}\,}
\def\simle{\,\hbox{\hbox{$ < $}\kern -0.8em \lower 1.0ex\hbox{$\sim$}}\,}
\def\wig#1{\mathrel{\hbox{\hbox to 0pt{%
          \lower.5ex\hbox{$\sim$}\hss}\raise.4ex\hbox{$#1$}}}}
\newcommand\etal{{\it et al.}}

\title{\Large \bf The Galactic disk mass-budget:\\ II. Brown dwarf mass-function and density.}

\author{Gilles Chabrier}
\affil{
Ecole Normale Superieure de Lyon,\\
C.R.A.L. (UMR 5574 CNRS),
69364 Lyon Cedex 07, France\\
and\\
Visiting Miller Professor, Dpt. of Astronomy, University of Berkeley,\\
Berkeley, CA 94720}

\authoremail{chabrier@ens-lyon.fr}

\begin{center}
\underline{Submitted\ to}$: {\sl Astrophysical Journal}$

\bigskip
\end{center}
\bigskip

\begin{abstract}

In this paper, we extend the calculations conducted previously in the stellar regime (Chabrier, 2001) to determine the brown dwarf initial mass function in
the Galactic disk. We perform Monte Carlo calculations taking into account
the brown dwarf formation rate, spatial distribution and
binary fraction. Comparison with existing surveys seems to exclude a power-law
mass function as steep as the one determined in the stellar regime below 1 $\msol$, i.e. $dn/dm\propto m^{-1.5}$, and tends to favor a more flatish behaviour. Although a power-law mass function in the substellar regime can not
be excluded by present day observational constraints, a form $dn/dm\propto m^{-1}$, i.e. $dn/d\log m=constant$, seems to be an upper limit.
Comparison with methane-dwarf detections tends to favor an eventually decreasing
form like the lognormal or the more general exponential distributions determined in the previous paper. We calculate predicting brown dwarf counts in near-infrared color diagrams and brown dwarf discovery functions for various types of mass functions and formation rates, and for different binary distributions. Based on these diagnostics, future large deep field surveys
should be able to determine more precisely the brown dwarf mass function and to
provide information about the formation of star-like objects - stars and brown dwarfs - along the Galactic history. These calculations
yield the presently most accurate determination of the brown dwarf census in the Galactic disk. The brown dwarf number density is comparable
to the stellar one, $n_{BD}\simeq
n_\star\simeq 0.1$ pc$^{-3}$, showing that the star formation process in the disk
extends well into the substellar regime. The corresponding brown dwarf mass density, however, represents only about
10\% of the stellar contribution, i.e. $\rho_{BD}\simle 5.0\times 10^{-3}\,\mvol$. Adding up the local stellar density determined in the previous
paper, we obtain the density of star-like objects, stars and brown dwarfs, in
the solar neighborhood $\rho_\odot \approx 5.0\times 10^{-2}\,\mvol$.
\end{abstract}
\bigskip



\section{Introduction}

Brown dwarfs (BD) - objects not massive enough to sustain proton fusion in their core - were once thought to be the
most favorable candidates to make up for the Galactic missing mass. Indeed their luminosity is orders of magnitude fainter than the Sun and a Salpeter-like mass function (MF) extending into the BD domain would yield a large
mass-to-light ratio and thus a significant amount of baryonic dark matter in the Galactic disk or halo.
Although this scenario is now clearly excluded, a proper
census of the number of BDs in the Galaxy has significant implications for our understanding of how star-like
objects form and for an accurate determination of the contribution of substellar objects to the Galaxy mass
budget, more particularly in the disk. Since by definition BDs never reach
thermal equilibrium and formed at several epochs during the evolution of the Galaxy, this task requires not only the determination of the BD initial mass function
(BDIMF), i.e. the extension of the stellar IMF below the hydrogen-burning limit,
but also of the BD birthrate. The determination of these quantities from today BD observable signatures
implies the knowledge of BD evolution, i.e. accurate relations between mass, age, and observable quantities such as magnitudes and colors.

Observations of BDs in the solar neighborhood have progressed at a rapid pace within the
past few years.
The DENIS (Delfosse et al., 1999) and 2MASS (Kirkpatrick et al. 1999; Burgasser et al. 1999) surveys, which have covered initially
areas of several hundred square degrees and are complete to $K\simeq 13.5$ and $K_S\simeq 14.5$, respectively, have first revealed about
20 field L-dwarfs and 4 field methane-BDs. This yields an L-dwarf number-density $n_L\approx 0.03$ (sq deg)$^{-1}$ for $K_S<14.5$ and a methane-BD (also denominated T-dwarfs) number-density $n_{CH_4}\approx 0.002$ (sq deg)$^{-1}$ for $J<16$, with large uncertainties. These determinations have improved with subsequent L-dwarf and T-dwarf discoveries with the Sloan survey (Strauss et al., 1999; Tsvetanov et al., 2000; Leggett et al., 2000), the NTT Deep Field (Cuby et al., 1999) and with the extension of the 2MASS survey wich covers now about 10\% of the sky with a color-selection criterion $J-K_s\ge 1.3$ (Kirkpatrick et al., 2000). These surveys bring the number of discovered L- and T-dwarfs to over 100 and about 10, respectively, almost half of the L-dwarfs and most of the T-dwarfs being suspected to lie within $\sim 25$ pc from the Sun. Even though statistics is
still small, these numbers provide important observational constraints
on the present-day BD number-density in the Galactic disk.

A first attempt to determine the BD density and BDIMF has been conducted recently by Reid et al. (1999). Although
these calculations certainly point the way, they rely on BD models primarily devoted to the description of Gl-229B like objects (i.e. methane-BDs) (Burrows et al., 1997) and thus
do not apply to the characteristics, spectroscopic or photometric signature,
of L-dwarfs. Moreover, they rely on
dubious color-$\te$-bolometric correction relations, and thus
suffer from inconsistencies in the mass-age-magnitude-color
relations. At last, these calculations assume one single form for the IMF, namely a power-law function.
It is the aim of the present paper to improve upon these calculations by using consistent BD evolutionary
calculations and by considering several possibilities for the BD Galactic history, mass function, birthrate and binary frequency. The IMF in the stellar regime has been determined in a
previous paper (Chabrier, 2001, hereafter Paper I) down to the vicinity of the H-burning limit, providing the normalization at this limit to extend into the BD domain. This stellar IMF has been shown to be adequately
fitted by three
different functional forms over the entire stellar mass-range.
These functional forms, however, behave quite differently in the BD regime. While the power-law function keeps
rising monotonically, the two other forms - lognormal or exponential -  slowly decline in the substellar domain below a characteristic mass.
We thus expect the density of BDs to differ appreciably, depending on the form of the adopted BDIMF, at least between a power-law as the one determined in Paper I and a lognormal or exponential form.
The present calculations should help distinguishing between these two generic forms in the low-mass domain, power-law or lognormal,
by comparison with available BD observations. For a sake of completeness, we also
conduct calculations with a power-law form with $\alpha=1$, i.e. $dn/d\log m =$constant, in the substellar regime. We calculate the predicted
BD discovery function (BDDF) and BD counts associated with these different IMFs for large-field surveys
at faint magnitude. These predictions will serve as a guide to analyze or to design future surveys aimed at determining
the stellar and brown dwarf local densities and Galactic history.

In \S2, we briefly comment on the BD models used in the present calculations. In \S3, we outline the
calculations and describe the BD mass, age, spatial and binary probability distributions. The
predicted BD counts and BDDF, and comparisons with existing surveys, are presented in \S 4. The derived BD number- and mass-densities in the Galactic disk
are presented in \S5, which yields the presently most accurate determination
of the stellar+BD density in the solar neighborhood. Section 6 is devoted to the conclusion.

\section{Brown dwarf evolutionary models}

\subsection{From L-dwarfs to T-dwarfs}

Brown dwarfs are now identified as belonging to two main spectral types, the so-called "L-dwarf" and "methane-dwarf" types, the latter being also denominated "T-dwarf". The L-dwarfs are characterized by very red near infrared colors, with $J-K\simgr 1.0$, a consequence of the formation of grains near their photosphere, whereas the second population exhibits bluer near infrared colors, with $J-K\simle 0.5$, due to the dominant absorption of methane overtone bands in the 1.0 to 2.5 $\mu$m range (H and K bands) (see e.g. Figure 9 of Kirkpatrick et al., 2000). The effective temperature range characteristic of the L-dwarf population remains ill-determined. Basri et al. (2000), using a $\te$-$Sp$ classification based on the analysis of CsI and RbI lines with a preliminary set of synthetic spectra of Allard et al. (2001), obtain a range $\te \approx 1600-2200$ K, consistent with the values obtained with evolutionary models in color-magnitude
diagrams (Chabrier et al., 2000a), whereas the more empirical $\te$-$Sp$ relation derived by Kirkpatrick et al. (1999) yields a cooler temperature range $\te \approx 1300-2000$ K. New analysis of Keck HIRES spectra with improved dusty-atmosphere models are in reasonable agreement with the Basri et al. (2000) determination (Schweitzer et al., 2001). Although it is still premature to settle for one of these two determinations, recent observation of the onset of methane absorption in the L-dwarf 2M1507 (Noll et al., 2000) tends to support the former scale. Indeed the weakness of the $\nu_3$ methane band in this object, and in 2M0825, suggests a temperature significantly higher than the CH$_4$-CO equilibrium temperature $T\simeq 1200-1400$ K under BD atmosphere conditions (photospheric pressure $P_{ph}\sim 3-10$ bars) (Tsuji et al 1995; Fegley \& Lodders, 1996), a value which would result from the Kirkpatrick et al. (1999) determination.

As argued by Kirkpatrick et al. (1999), their temperature scale yields a temperature gap of $\sim 350$ K between the faintest observed L-dwarf Gl584C and the T-dwarf Gl229B, for a $\sim 0.4$ magnitude difference, whereas the Basri et al. (2000) scale corresponds to about twice this temperature gap. Although such an argument is certainly relevant, the quantitative result must be considered
with caution for several reasons. First of all, the temperature gap relies on $\te$ estimated from the spectral type, i.e. the energy spectral distribution only. However, grain formation in L-dwarf atmospheres yields a severe backwarming effect. Therefore, the difference in the {\it thermal structure} of the atmosphere between an L-dwarf and a T-dwarf atmosphere, where most grains are suspected to form or settle much below the photosphere, is already significant (see e.g. Figure 1 of Chabrier et al., 2000a). It would be no surprise that this grain settling effect yields by itself a considerable decrease in $\te$ between (dusty) L-dwarfs and (almost grainless) Gl229B-like objects. The quantification of this effect must await atmosphere models including grain sedimentation. The other possible flaw in the Kirkpatrick et al. (1999) $\te$ determination stems from their estimate of the absolute magnitude of Gl584C. This latter is determined from a $M_J$ or $M_K$ vs $Sp$ relationship based on the Kirkpatrick et al. (1999) L-dwarf spectral classification. Until this classification is proven unambiguously to be the correct one, the magnitude determination must be taken with great caution. At last, these authors assume the same mass, $\sim 0.045\,\msol$, and the same age, $\sim 0.5$ Gyr, for Gl584C and Gl229B, two rather arbitrary assumptions. The M$_J$ magnitude of L-dwarfs with mass $m\sim 0.05\,\msol$ varies by about 6 magnitudes between 0.1 and 1 Gyr, not mentioning the strong variation of magnitude with mass (see e.g. Chabrier et al., 2000a).

The more recent $\te$-Sp relationship for early
L-dwarfs of Schweitzer et al. (2001) and the spectroscopic analysis of Leggett et al. (2001) suggest that dust settling in the
atmosphere of L-dwarfs should start around $\te\sim 1700$-1800 K, and that
the limit case illustrated by the so-called "dusty" atmosphere models should no
longer be used below about this temperature. We will respect this constraint in the present
models (see \S 2.2).
A much more robust, severe constraint for the transition from L-dwarfs to T-dwarfs is the null detection of objects with $J-K\simgr 2.1$ in all surveys, covering about 10\% of the sky, whereas objects fainter than the ones at this limit have been discovered, all with T-dwarf characteristic $J-K\sim 0$ (Kirkpatrick et al., 2000).

More recently, objects with weak CH$_4$ absorption features in the H and K bands have been discovered with the Sloan survey (Leggett et al., 2000) with intermediate colors $0.5\simle J-K\simle 1.0$ and $0\simle H-K\simle 0.5$ and have been identified as early T-dwarfs.

\subsection{The models}

The BD evolutionary models used in the present calculations are described in Chabrier et al. (2000a) and
in Chabrier \& Baraffe (2000). The first set of models are the so-called "dusty" models, based on the recent Allard et al. (2001) atmosphere
models which include grain formation in
the atmosphere equation-of-state (EOS) and grain opacity in the radiative transfer equation.
These models successfully reproduce the
observed colors and magnitudes of the afore-mentioned L-dwarfs near the bottom of the main sequence (Chabrier et al., 2000a), as well as their spectral energy distribution (Leggett et al., 2001; Schweitzer et al., 2001). The second set of models,
so-called "cond" models, include grain formation in the atmosphere EOS, thus taking into account the corresponding
element depletion, but ignore the grain opacity. This case mimics a rapid settling of grains
below the photosphere and reproduces reasonably well the colors and magnitudes of Gliese229B and methane BDs
(see Chabrier et al., 2000a and figure 13 of Chabrier \& Baraffe, 2000).

In order to compute a complete evolutionary sequence over the entire BD range from the hydrogen-burning limit
$m\simeq 0.07\,\msol$ down to a Jupiter mass 0.001 $\msol$, we have used the dusty-models for $\te \ge 1700$ K and the cond-models
for $\te \le 1350$ K, based on the arguments discussed in \S 2.1, and we have  smoothly interpolated between these two limits for objects with effective temperatures
in-between.

These models provide consistent relationships between mass, age, effective temperature, colors and magnitudes
for the disk BD population, avoiding dubious transformations of M$_{bol}$ or $\te$ into observable quantities. Although still far from an accurate description of the various phenomena at play in L-dwarf and T-dwarf atmospheres, and thus from a robust description of their observational signatures for a given mass and age,
they reproduce quite well, as mentioned above, the only robust observational constraints on these objects, namely magnitudes and colors,
in the two limit cases of hot, dust-dominated
L-dwarfs and Gl229B-like or cooler methane-dwarfs, respectively. In particular in near-infrared colors, the relevant spectral domain for BD detection. Indeed, the effective temperature itself is not the issue in the present calculations; it is used only to interpolate between L-dwarfs and methane-dwarfs. The main uncertainty occurs obviously for the objects in this interpolated region, where there is presently only one object with determined parallax and thus
absolute magnitude (Els et al., 2001). The present models reproduce quite well
this important constraint both in $\mj$ vs $J$-$H$ and $\mj$ vs $J$-$K$ color-magnitude diagrams, respectively within and at the limit of the observational error bar,
for an age $t\sim$1 Gyr, the expected age of the system. This brings confidence
in our interpolation procedure. Note that this domain of interpolation covers only about one magnitude in
$J$ (Chabrier et al., 2000a; Chabrier \& Baraffe, 2000). As shown in \S4, objects in this region correspond to a restricted combination of masses and ages, namely young low-mass BDs or older massive BDs, and represent only a minor fraction of the Galactic BD population. Given these facts, the present calculations should yield reasonably accurate determinations of the BD color and magnitudes in term of their age and mass.

\section{The calculations}

We use Monte-Carlo simulations to generate a sample of BDs with known mass, age and distance.
The luminosity, effective temperature and colors are then given by the models described in \S2.
We also consider the possibility of BD binary systems, with various frequencies and mass ratios.
This yields the Luminosity Function (LF) of the systems and of the resolved objects. The various stages of the calculations
are outlined below.

\subsection{Total number of objects in the simulations}

The total number of objects $N_{tot}$ in the simulated volume is:

\begin{eqnarray}
N_{tot}=\Omega \times \Bigl( \int_0^{d_{max}} n(r)\,r^2\,dr \Bigr) \times
\Bigl( \int_{m_{inf}}^{m_{sup}} \xi(m)\,dm \Bigr)
\label{eqntot}
\end{eqnarray}

\noindent where $\Omega$ is the field of view, $n(r)$ is the spatial distribution
and $\xi(m)={dN\over dm}$ denotes the initial mass function, i.e. the number of
objects ever formed in the mass-interval $[ m, m+dm ]$.
The first term in brackets in eqn.(\ref{eqntot}) is the volume integration.
We use a standard double-exponential disk
for the spatial density distribution:

\begin{eqnarray}
n\bigl ( z,R \bigr )= 1.0\times\,e^{-({R-R_\odot\over L})-{{\| z\|}\over h}}
\label{eqnrho}
\end{eqnarray}

\noindent where $z(l,b)$ and $R(l,b)$ are the galactocentric cylindrical coordinates for longitude $l$ and latitude $b$,
$L\simeq 2500$ pc is the scale length, $h\simeq 250$ pc the scale height (Haywood, Robin \& Cr\'ez\'e, 1997) and
$R_\odot=8.5$ kpc is the Sun galactocentric distance.

Note that the number-density normalization in eqn.(\ref{eqnrho}) is set up to 1 pc$^{-3}$; the mass-density normalization
in the solar neighborhood is fixed by the observationaly-determined value of
the IMF at a given mass (eqn.(\ref{eqnbdmf}) below) and thus by the integral of this IMF over the stellar+substellar mass range (see \S 3.3 and \S 5).
For the simulations, we chose a characteristic distance limit
for BD detection $d_{max}$ of a few kpc in eqn.(\ref{eqntot}), so that the number of
objects in the simulations is $\sim 10^6$. Methane-BDs have been
detected at about 100 pc with the NTT Deep Field survey (Cuby et al., 1999). The maximum mass for BDs, i.e. the minimum mass for stable hydrogen-fusion is $m_{sup}\simeq 0.072\,\msol$
(Chabrier \& Baraffe, 1997) and the minimum mass is chosen to be $m_{inf}=0.001\,\msol$, about a Jupiter mass.
For illustration,
for a limit magnitude $J_{lim}\sim 22$, a distance limit d=100 pc ($M_J\simeq 17$) corresponds to
the limit of detection of methane-BDs
$m\simeq 0.001\,\msol$ at $t\simeq 10 ^6$ yr,
$m\simeq 0.01\,\msol$ at $t\simeq 3\times 10 ^8$ yr,
$m\simeq 0.02\,\msol$ at $t\simeq 10 ^9$ yr
and $m\simgr 0.05\,\msol$ at $t\simeq 10 ^{10}$ yr (Chabrier \& Baraffe, 2000).

\subsection{The brown dwarf initial mass function}

The BDIMF is the extension in the substellar regime of the stellar IMF determined in Paper I. Unlike stars, which eventually evolve off the main sequence stage, BDs have
unlimited lifetimes so that all BDs ever formed in the Galaxy still exist today,
regardless on when they were formed, and the present day BD mass function
is the BD initial mass function.
We conducted the calculations with three different forms of IMFs,
namely the power-law IMF labeled IMF1 in Paper I:

\begin{eqnarray}
\xi(m)={dn\over dm}=A\,m^{-\alpha}
\label{eqnimf1}
\end{eqnarray}

\noindent where $n$ is the number-density of objects, $A=0.019$ $\msol^{-1}pc^{-3}$, $\alpha=1.55$ (see Paper I),

\noindent the lognormal IMF labeled IMF2 in Paper I:

\begin{eqnarray}
\xi(\log \,m)={dn\over d\log \, m}=A\,exp\{-{(\log \, m\,\,-\,\,\log \, m_0)^2\over 2\,\sigma^2}\}
\label{eqnimf2}
\end{eqnarray}

\noindent with $A=0.141$ pc$^{-3}$, $m_0=0.1\,\msol$ and $\sigma=0.627$ (see Paper I).

As mentioned in Paper I, the lognormal form (\ref{eqnimf2}) and the exponential form (IMF3, eqn.(8) of Paper I) are
barely distinguishable in the BD domain and yield very similar results.

For a sake of completeness, we have also performed calculations with a power-law with
a shallower slope $\alpha =1.0$ in the BD domain, as suggested in previous calculations (Reid et al., 1999). To insure the continuity of the MF, the normalization is fixed at 0.1 $\msol$ at the value given by IMF1, $dn/dm (0.1\,\msol)=0.674\,\msol^{-1}$pc$^{-3}$. This IMF, hereafter denoted IMF4, thus reads:

\begin{eqnarray}
\xi(m)={dn\over dm}=A\,({m\over 0.1})^{-1.55}\,\,\,\,\,\, m\ge 0.1\,\msol \nonumber\\
\,\,\,\,\,\,\,\,\,=A\,({m\over 0.1})^{-1.0}\,\,\,\,\,\, m\le 0.1\,\msol
\label{eqnimf4}
\end{eqnarray}

\noindent with $A=0.674$ $\msol^{-1}pc^{-3}$
 
The total number-density of objects in the mass-interval $[m_{inf},m_{sup}]$ is

\begin{eqnarray} n=\int_{m_{inf}}^{m_{sup}}\xi(m)\,dm=\int_{\log\,m_{inf}}^{\log\,m_{sup}} \xi(\log\,m)\,d\,\log\,m
\label{eqnnumber}
\end{eqnarray}

We suppose in the present calculations that the IMF does
not depend on time and remained constant along the evolution of the Galactic disk.

\subsection{The formation rate}

We consider two types of star (more generally star-like object) formation rates (SFR), namely:

\indent - a constant SFR:

\begin{eqnarray}
b(t)=b_0=constant
\label{eqnsfr1}
\end{eqnarray}

\indent - a time-decreasing exponential SFR:

\begin{eqnarray}
b(t)=b_0\,e^{-{t-t_{inf}\over \tau}}
\label{eqnsfr2}
\end{eqnarray}

\noindent with an e-folding time $\tau=5$ Gyr (Miller \& Scalo, 1979).

The values of $b_0$ obey the normalization condition
${1\over \tau_G} \int_{t_{inf}}^{\tau_G} b(t) dt=1$, where $\tau_G\approx 10$ Gyr is the age
of the Galactic disk and $t_{inf}<< \tau_G$ (Miller \& Scalo, 1979; Scalo, 1986). The (roughly) constant SFR is the most favored solution for the disk history (see e.g. Scalo, 1986) but comparison with an exponential
SFR is instructive, as shown below.

As mentioned previously, the normalization of the BDIMF in the present calculations is imposed by the number of objects observed today at a given
mass, e.g. at the H-burning limit $m\approx$0.07 $\msol$. The total number-density of star-like objects
(stars plus BDs)
ever formed in the Galactic disk is given by the integral of the creation function (Scalo, 1986):

\begin{eqnarray}
n_{tot}(t=\tau_G)={1\over \tau_G}\int_0^{\tau_G}b(t)dt \times \int_{m_{inf}}^\infty \xi(m)dm
=\int_{m_{inf}}^\infty \xi(m)dm
\end{eqnarray}

\noindent where the afore-mentioned normalization condition on the SFR
has been applied and where we assume separability between the SFR and the IMF, so that the total density of objects per unit mass at the H-burning limit today is:

\begin{eqnarray}
{dn_{tot}\over dm}(m=0.07\,\msol)=\xi(m=0.07)
\label{eqnbdmf}
\end{eqnarray}

This value corresponds to 1.17 $\msol^{-1}$ pc$^{-3}$ for IMF1
(eqn.(\ref{eqnimf1})), 0.85 $\msol^{-1}$ pc$^{-3}$ for IMF2 (eqn.(\ref{eqnimf2})) or IMF3 (see Paper I),
and 0.96 $\msol^{-1}$ pc$^{-3}$ for IMF4 (eqn.(\ref{eqnimf4})), respectively.

\subsection{The probability distributions}

For each object $i=1,N_{tot}$, the Monte Carlo technique is used to determine:

\indent $\bullet$ its age ($\tau_G-t_i$) with a probability law $P(t)=\tau_G^{-1}\times\int_{t_{inf}}^t b(t^\prime)\,dt^\prime$ where $b(t)$ is
the probability density given by either $b(t)=constant$ or $b(t)=e^{-{t-t_{inf}\over \tau}}$,
with the normalization condition $\tau_G^{-1}\times\int_{t_{inf}}^{t_{sup}} b(t)\,dt=1$.
The lower and upper limits are chosen as $t_{inf}=10^6$ yr, $t_{sup}=\tau_G$.

\indent $\bullet$ its distance $r_i$ with a probability law $P(r)=\int_0^r p(r^\prime)\,dr^\prime$
where $p(r)$ is the probability density $p(r)=r^2\,n(r)$,
for $r=0\rightarrow d_{max}$, and $n(r)$ is given by eqn.(\ref{eqnrho}).

\indent $\bullet$ its mass $m_i$ with a probability law $P(m)=\int_{m_{inf}}^m p(m^\prime)\,dm^\prime$
where $p(m)$ is the probability density given either by the uniform distributions (\ref{eqnimf1}) or (\ref{eqnimf4}), or by the gaussian distribution (\ref{eqnimf2}),
with the normalization condition $\int_{m_{inf}}^{m_{sup}} p(m)\,dm=1$.

We consider also the possibility for a uniformly distributed
fraction $X_{BD}$ of BDs to have another BD companion. For the mass ratio between the secondary
and the primary $q(m)={m_2\over m_1} \le 1$, we have conducted the calculations with two different
distributions, namely:

\indent - a normal form:
\begin{eqnarray}
P(q)=e^{-{(q-m_c)^2\over 2\,\sigma^2}}
\label{eqnpq1}
\end{eqnarray}

\noindent where $m_c=0.23$ and $ \sigma =
  0.42 $ are taken from Duquennoy \& Mayor (1991).

\indent - a uniform distribution:
\begin{eqnarray}
P(q)=constant
\label{eqnpq2}
\end{eqnarray}

\section{The results}

The probability distributions described in \S3.4 yield for each object $i=1,N_{tot}$ its mass $m_i$, age $\tau_G-t_i$,
distance $d_i$. The properties $T_{{eff}_i}$, $\log \, g_i$, absolute magnitude $M_i$, apparent
magnitude $m_{{ap}_i}$, and colors are given for each $(m_i,t_i)$ by the models
described in \S2.2. The afore-mentioned probability distributions give also
the number $N_{bin}=X_{BD}\times N_{tot}$ of BDs with a companion of
mass $m_{i_2}=q(m).m_{i_1}$.
The apparent magnitudes of the
{\it systems} in a filter $\lambda$ are derived from the values of the two unresolved companions:

\begin{eqnarray}
m_{{ap}_\lambda}(sys)=
m_{{ap}_\lambda}(m_1)\,-\,2.5\,\log \{1+10^{({m_{{ap}_\lambda}(m_1)-m_{{ap}_\lambda}(m_2)\over 2.5})}\}
\label{eqnsys}
\end{eqnarray}

This yields the so-called systemic distribution.

For a survey with a magnitude-limit $m_{ap_{lim}}$, only the objects with $m_{{ap}_i}\le m_{ap_{lim}}$ are
considered in the results of the simulation.

\subsection {Comparison with existing surveys}

In order to test the validity of our calculations, we first confront the theoretically predicted numbers with existing surveys. Such results are given in Table I.
The DENIS survey (Delfosse et al., 1999) has a field-of-view $\Omega = 240$ sq.deg, a magnitude limit $K_{lim}\simeq 13.5$. It is expected to
detect L-dwarfs up to a distance $d_{max}\simeq 30$-35 pc, although the
exact volume probed by BD surveys is not well determined, since $d_{max}$ strongly
varies with the mass and the age of each BD. The 2MASS
survey (Kirkpatrick et al., 1999; Burgasser et al., 1999) had originally $\Omega=371$ sq.deg.,
$K_{S_{lim}}\simeq 14.5$ ($K_{lim}\simeq 15$) (see Table 3 of Kirkpatrick et al., 1999) and should detect L-dwarfs up to about 50 pc, with a color selection criterion
$J-K> 1.3$. It has been extended recently to about 10\% of the sky, i.e. $\Omega\simeq 4000$ sq.deg.,
$K_{lim}\simeq 15$ (Kirkpatrick et al., 2000; Burgasser, 2000).
Almost half of the detected L-dwarfs are suspected to lie
within 25 pc from the Sun, although the distances are estimated from the
Kirkpatrick et al. (1999) absolute magnitude vs spectral type relation and
must be considered with caution. Only 17 L-dwarfs have measured parallaxes.
This survey, however,
is likely to be substantially incomplete (D. Kirkpatrick, private communication) so that the number of L-dwarfs detected in the whole survey
given in Table 1 is certainly largely underestimated. The
incompleteness is clearly illustrated by the comparison between observation
and theory or by a simple scaling
between the two 2MASS fields. From our calculations, most of these L-dwarfs are
found to lie within about 50 pc from the Sun.

Other surveys like the Sloan Digitized Sky Survey (Strauss et al., 1999;
Tsvetanov et al., 1999; Leggett et al., 2000) and the ESO NTT survey (Cuby et al., 1999) have
also discovered several L-dwarfs and methane-dwarfs in the solar neighborhood. Note that for methane-dwarfs, the color selection used by 2MASS, J-H$\le 0.3$,
H-K$\le 0.3$,
eliminates early (hot) methane-dwarfs, in the L/T BD transition domain, as the ones
discovered by SDSS, in spite of the smaller volume probed by this survey. The majority of the T-dwarfs
identified by 2MASS and SDSS have $J\sim$ 15-16, making then detectable up to
about 12 pc for Gl229B-like objects ($\te\approx 1000$ K) (see Chabrier et al., 2000a).
The number of objects found by these different surveys are
given in Table I.

For this comparison we consider only the case of a constant SFR (eqn.(\ref{eqnsfr1})). The case of a time-decreasing SFR will be examined in \S4.2.4. A number of BD
binary systems have been resolved observationally in the afore-mentioned surveys. For the conditions of these surveys and a
BD binary fraction $X_{BD}=0.3$, however,
the numbers of theoretically predicted detections of systems or resolved objects are not significantly different ($\sim 10\%$). For L-dwarf detections, we used $m_{sup}=0.08\,\msol$. Indeed, very-low-mass hydrogen-burning stars,
with $0.072\,\msol< m\simle 0.08\,\msol$ and $t\simgr 1$ Gyr reach effective temperatures characteristic of the L-dwarf type (see e.g. Fig. 16 of Chabrier \& Baraffe 2000
or Tables 3-5 of Chabrier et al. 2000a).

The comparison between theory and observations in Table I
should be taken with caution. Statistics
is still small, models still uncertain, the true BD binary fraction presently
unknown and the exact limit magnitudes of the survey not well established. In spite of these uncertainties, some information can be drawn. The L-dwarf surveys seem to favor
the lognormal IMF2 or the shallower power-law IMF4, the IMF1 predicting a substantially larger number of objects
than observed. However, as illustrated by the
2MASSII survey, large fields
are necessary to really distinguish between these forms if only L-dwarfs are considered. The T-dwarf detections, however, seem to exclude
the IMF1, which overestimates significantly the number of T-dwarfs
detected by 2MASS.
The lognormal form
IMF2 yields also a good agreement with T-dwarf observations,
whereas the number of T-dwarfs predicted with IMF4 appears to be overestimated. A more robust
determination between these two latter forms, however, needs larger statistics.

\subsection {Brown dwarf mass, age and temperature distributions}

In this section we conduct predictive calculations for large, deep field surveys, which should yield enough statistics to determine more precisely the shape of the BDIMF. We also examine the predicted
distributions of BD masses, ages and effective temperatures as a function of magnitudes and colors. We define the following reference model:

\begin{eqnarray}
\nonumber
X_{BD}=0.30\,\,;\,\, P(q)=constant\,\,;\,\,
\label{eqncondn}
\end{eqnarray}

and reference survey:

\begin{eqnarray}
\nonumber
J_{lim}=22\,\,;\,\, \Omega=10\,{\rm sq.deg.}\,\,;\,\,(l,b)=(130^{o},25^{o})
\label{eqncondn}
\end{eqnarray}

It is instructive to examine the expected mass-, age- and $\te$-distributions of the BDs as a function of near infrared magnitudes or colors for the conditions mentioned above. Figure \ref{chab2_fig_mass.ps} displays the distribution
of BDs ($m\le 0.072\,\msol$)
for IMF1, IMF2 and IMF4.
As expected, the IMF1 predicts a significantly larger number of BDs, in particular of very-low-mass BDs, below
the deuterium-burning minimum mass $m_D\simeq 0.012\,\msol$. The IMF1 predicts about 200 of these objects in our reference survey whereas IMF4 predicts about 60 and only a few
are predicted to be visible with IMF2. This provides an interesting diagnostic on the
shape of the BDIMF. This requires, however, the spectroscopic observation of the presence of
deuterium at faint magnitude, a delicate, although achievable task (Chabrier et al., 2000b). Note that young ($\simle 10^7$ yr) BDs above the deuterium-burning minimum mass have not burned their deuterium yet (Chabrier et al., 2000b, Figure 1), but the relative contribution of such young objects in the field is statistically negligible
(see below).

Figure \ref{chab2_fig_age_col.ps} displays the BD age-distribution in $J$-$K$ colors obtained with IMF1 and IMF2.
 As seen in the figure,
there is a noticeable accumulation of young L-dwarfs, with $t\simle 10^8$ yr, redward of $J$-$K\simgr 0.8$.
This reflects the cumulative contributions
of massive ($\simgr 0.05\,\msol$) BDs and of very young
($t\simle 10^7$ yr) BDs near the deuterium burning minimum mass ($m\simle 0.03\,\msol$).
Deuterium-burning
keeps these objects hot enough ($\te > 2000\,{\rm K}$) to exhibit late-M and L-dwarf colors (see Chabrier et al., 2000b, Figure 1 and Chabrier et al., 2000a, Figure 6). More massive BDs will burn their deuterium content at earlier ages and the number of such very young objects is statistically insignificant.
The various loops redward of $J$-$K\sim 1.0$ at various ages reflect BD
cooling for different masses.
The other statistically dominant BD population in the color
diagram lies blueward
of $J$-$K\sim 0.8$, which corresponds to $\te\simle 1400$ K (see Figure \ref{chab2_fig_teff.ps}).
This is the consequence of both
the contribution of lower mass BDs
(see Figure \ref{chab2_fig_mass.ps}) and of BD cooling, since, as seen in
Figure \ref{chab2_fig_age_col.ps}, the BD distribution
is dominated by objects with an age $t > 10^9$ yr .
As a consequence of these two statistically favorably populated regions in the color diagram,
respectively redward of $J$-$K\sim 0.9$ and blueward of $J$-$K\sim 0.5$, there
is a relative scarcity of objects in the region in-between. This is illustrated
in Figure \ref{chab2_fig_color.ps} which displays the BD distribution for the two IMFs as a function of $J$-$K$ color. For samples with
low statistics, as for presently existing surveys,
 this scarcity can lead to a gap, with no object discovered in this
color interval. The age-distribution obtained with IMF1 exhibits the same qualitative behaviour as with IMF2, except for a larger number of young ($<10^8$ yr) T-dwarfs ($J$-$K< 0.6$), as expected from the larger number of very-low mass BDs (see Figure \ref{chab2_fig_mass.ps}).

Figure \ref{chab2_fig_teff.ps} displays the $\te$ vs $J$-$K$ relation for the
predicted BD distribution. Note in passing the presence of 
hot ($\te \ge 2500$ K) genuine BDs ($m\le 0.07\,\msol$). These objects
have a V-magnitude $M_V\simle 16$.
Therefore, some of the faintest objects identified as very-low-mass stars in the local sample may in fact be young, bright BDs, yielding a BD contamination of the last bin(s) of the 5-pc nearby LF, as suggested in Paper I.

\subsection {Predicted brown dwarf discovery function}

In this section we calculate the predicted BDDF for the reference
survey
mentioned in \S4.2\footnote{In this paper, we stick to the following definitions: the luminosity function is the number of objects per absolute
magnitude interval whereas the discovery function is the predicted number of objects to be found by a survey of given area and limited magnitude, as considered in the present section}, and we examine the dependence of this DF upon the
different inputs for BD Galactic history, IMF, SFR and binary frequency.

\subsubsection {Effect of the IMF}

Figure \ref{chab2_figIMF_col.ps} compares the systemic
apparent BDDF in the J-band $\Phi_{J_{sys}}$
obtained with the BDIMFs IMF1, IMF2 and IMF4, respectively. The sudden increase of the
BDDF around $J\sim 20$-21 stems form the onset of the T-dwarf contribution,
as illustrated by the dotted and dashed lines, respectively, for IMF1 and IMF2.
As expected from the larger number of predicted low-mass BDs, the number
of faint objects predicted with the steeply rising IMF1 is significantly larger than with
the two other forms. As expected from Figure \ref{chab2_fig_mass.ps}, the number
of L-dwarfs ($J$-$K\simgr 0.8$) predicted with IMF2 and IMF4 is comparable, but this latter form predicts almost twice as many T-dwarfs.
The L-dwarf to T-dwarf ratio amounts to $n_L/n_T\approx 2$ with IMF1,
$n_L/n_T\approx 2.5$ with IMF4 and
$n_L/n_T\approx 3$ with IMF2. Large field photometric surveys at faint
magnitude should thus be able to provide important information on the shape of the BDIMF, simply
by looking at the faint end of the distribution and at the $n_L/n_T$ ratio.

\subsubsection {Effect of the SFR}

Figures \ref{chab2_figSFR_col.ps} compares the systemic
 apparent DFs $\Phi_{J_{sys}}$ obtained with IMF1 and IMF2,
 respectively, for two different
SFRs, namely a constant SFR (eqn.({\ref{eqnsfr1})) and an exponentially decreasing SFR (eqn.({\ref{eqnsfr2})).
As expected, the latter one yields a significantly smaller number of bright BDs since the majority of the
objects formed several billion years ago and have now dimmed to very faint magnitudes. The corresponding signature
on a color-magnitude diagram will be a smaller number of L-dwarfs. Indeed, BD counts done with the decreasing SFR for the conditions of the DENIS and 2MASS-I surveys (see
Table 1) predict 6 and 14 L-dwarfs, respectively, with IMF1, and 4 and 7 L-dwarfs
with IMF2. The number of T-dwarfs is also significantly diminished and the
combination of the exponentialy-decreasing SFR and the steep IMF1 yields a
creation function which brings
the predicted counts in
reasonable agreement with both the observed L-dwarf and T-dwarf detections. 

\subsubsection {Effect of binarity}

Although the frequency of BDs in multiple systems is presently very poorly determined, a few general trends emerge from the various existing searches:

- BDs seem to be very rare as close companions ($\simle 100$ AU) of nearby F, G and M stars, confirming the so-called "brown dwarf desert" at {\it small separation} (Halbwachs et al., 2000).
However BDs exist as wide companions ($\sim 100-4000$ AU) of solar-type stars
($\mv < 9.5$, $Sp=F$-$M0$, i.e. $m\simgr 0.5\,\msol$). No wide BD companion has been detected so far for lower mass stars. Although the fraction of such BDs is still very uncertain, Gizis et al. (2001) estimate this fraction to be $f_{BD}\approx 5-30\%$, which large uncertainties due to small statistics. From the stellar IMF determined in Paper I, the number density of solar-type stars with
$m\simgr 0.5\,\msol$ is $\sim 0.02$ pc$^{-3}$. With the afore-mentioned estimated value of $f_{BD}$, this yields a number density of BD as star companions $n_{BD\star} \approx 1.0$-$6.0\times 10^{-3}$ pc$^{-3}$ and an estimated mass density
$\rho_{BD\star} \le n_{BD\star}\times 0.07 \simle 0.5\times  10^{-3}\,\mvol$,
within at least a factor of 2 uncertainty. A better determination of
this fraction of BDs as wide companions of stars should emerge in the near future.

- About $\sim 20\%$ of L-dwarfs have an other L-dwarf companion. In contrast to
BDs companions of stars, these L-dwarf binaries have small ($\simle 10$ A.U.)
separations, whereas wider BD-BD systems are lacking (Reid et al., 2001). Several of the L-dwarf binaries discovered originally in the DENIS and 2MASS surveys
prove to be equal-luminosity, and thus equal-mass systems, although it is still premature to draw robust conclusions about the primary/secondary mass ratio.

Figure \ref{chab2_figBIN.ps} compares the systemic (unresolved) and the single (resolved) apparent J-band DFs
obtained
with $X_{BD}=0.30$ for three companion mass-distributions, namely
a uniform mass-ratio distribution (eqn.(\ref{eqnpq2}); long-dash line),
a gaussian distribution (eqn.(\ref{eqnpq1}); dotted line), and an equal-mass distribution $q(m)=1$ (short-dash line). For a sake of clarity, only calculations with IMF2 are displayed.
As seen in the figure, the correction on the systemic DF due to
unresolved companions amounts at most to a factor of $\sim 30\%$ at the faint end for the equal-mass case under the presently defined survey conditions.

\section{Brown dwarf number-density and mass-density}

The BD total number-density in the Galactic disk field $n_{BD}$ is given by
eqn.(\ref{eqnnumber}). We must add the contributions arising from BD companions in BD binary systems $n^\prime_{BD}=X_{BD}\times n_{BD}$ and from BD companions of main sequence stars $n_{BD\star}$ (\S4.3.3).

Similarly, the BD total mass-density in the Galactic disk is given, for an equal-mass distribution for the BD companions, by:

\begin{eqnarray}
\rho_{BD}=(1+X_{BD})\times \int_{0.001}^{0.07}\xi(m)mdm \, + \,
\rho_{BD\star}
\label{eqnbd2nbr}
\end{eqnarray}

Table 2 gives the corresponding numbers for the most favorable BDIMF IMF2, and
for the IMF4, obtained for a constant SFR for a BD-BD binary fraction $X_{BD}\approx 0.3$ with a mass-ratio $q(m)=1$ and a BD-star binary fraction $f_{BD}=$0.05-0.3. The corresponding BD surface densities are obtained for a disk scale height $h=250$ pc.

The present calculations and the ones derived in Paper I yield the presently
most accurate determination of the density in the solar neighborhood under the
form of stars and brown dwarfs. From Table 2 of Paper I and Table 2 of the present paper, one gets the local normalization for the disk $\rho_\odot\simeq (4.5+0.5)\times 10^{-2}\simeq 5.0\times 10^{-2}\,\mvol$,
$\Sigma_\odot\simeq (24.5+2.0)\simeq 26.5\,\msurf$, within about $\pm$10\%.

\section{Conclusion}

In this paper we have extended the stellar initial mass function determined in a
previous paper (Chabrier, 2001) into the brown dwarf domain. With this
mass function, we have performed Monte Carlo calculations, taking into
account the brown dwarf formation rate along the Galaxy history, the
brown dwarf spatial distribution in the disk and the possibility for
brown dwarfs to belong to binary systems with different frequencies and
mass ratio distributions. The calculations have been performed with the
two different functional forms for the mass function in the low-mass regime determined in Paper I, namely a power law $dN/d m \propto m^{-1.5}$, i.e.
 a linear log-log distribution, or a gaussian-type (lognormal)
 distribution $dN/d \log m \propto exp\{-{1\over 2\sigma^2}\log ({m\over m_0})^2\}$, both
 normalized to the observed stellar density at the bottom of the main sequence. We have also performed calculations with a shallower power-law MF below 0.1 $\msol$, $dN/d m \propto m^{-1.0}$, i.e $dN/d \log m= constant$. As shown in Paper I, the lognormal distribution is well reproduced in the low-mass domain by a more general exponential form.
Whereas the power law increases monotonically in the brown dwarf regime,
predicting an increasing number of objects per mass interval with decreasing mass, the lognormal and the exponential forms eventually turn down below a characteristic mass, about the hydrogen-burning
minimum mass in $dN/d \log m$ (see Figure 4 of Paper I), i.e. about the deuterium-burning minimum mass $m\simeq 0.012\,\msol$ in $dN/dm$.
 Comparison with existing L-dwarf and methane-dwarf detections tends to exclude
the continuation into the substellar regime of the power-law
 determined in the stellar domain, which predicts a number
 of BDs significantly larger than observed, at least for a constant star formation rate along the disk evolution. The shallower power-law yields
a number of L-dwarfs in agreement with the observations but seems to predict too
many T-dwarfs, although better statistics is needed to really nail down this
issue. These results show that a slope $\alpha=1$ is probably
about the upper limit for a power law fit of the IMF in the brown dwarf domain. Comparison with the observations seems to favor the
lognormal form, i.e. the general exponential form. As mentioned above, these results hold for a constant SFR. Degeneracy arises from various combinations
of the IMF and the SFR. Indeed a time-decreasing SFR combined with a steep BDIMF can lead to
BD counts comparable with the ones produced by a constant SFR and a shallower IMF. A nearly constant formation rate along the disk history, however, seems to give the best representation of
the high mass star distribution (see e.g. Miller \& Scalo, 1979) and of the
distribution of chromospheric activity in the local stellar population
(Henry, Soderblom \& Donahue, 1996).

 These results tend to support a flatening of the Galactic field
 stellar mass function $dN/d \log m$ around the H-burning limit. Although statistics is
 presently insufficient to accurately determine
the exact shape, i.e. the values of the various coefficients, of the mass function
 in the substellar regime, future large, deep field surveys should allow
 such a determination, in particular by comparing the relative contributions of the bright and faint parts of the brown dwarf luminosity function, and the photometrically identified ratio of methane dwarfs over L-dwarfs.

 Rapid progress both on the theoretical and the observational sides
 should quickly remove the remaining uncertainties in the present calculations. These latter present the first determination of
 a general, time-independent mass function, which seems to adequately describe the formation of star-like objects in the Galactic disk from about 100 $\msol$ down to about 10$^{-3}$ $\msol$, i.e. five orders of
 magnitude in mass, even though some uncertainty remains at very-low-masses between
a slowly rising power-law (with $\alpha<1$) and a lognormal form, because of presently limited statistics of field T-dwarf detections. If the general exponential form
 is confirmed, leading to a power-law form at large masses and a lognormal
distribution at low masses, it supports the
 suggestion that the formation of stars obeys a self-similar, statistically determined fragmentation process with no peculiar characteristic mass.
In that case, neither the hydrogen-burning nor the
deuterium-burning limits should play a peculiar role, but it is noteworthy that the number of objects per mass-interval in the lognormal or exponential distributions appear to turn down below about this latter limit.

Integration of this brown dwarf IMF yields the
presently most reliable determination of the BD census in the Galactic disk
and of its contribution to the Galactic disk mass budget. A significant source of uncertainty, however, stems from the yet undetermined contribution of BD companions of either other BDs or of stars at wide separations. Comparison with
the values determined in Chabrier (2001) for the stellar contribution shows
that the BD population in the disk is comparable to the stellar one, $n_{BD}\simeq
n_\star\simeq 0.1$ pc$^{-3}$, for the favored mass function IMF2,
so that the star formation process extends well into the substellar
regime. The BD mass contribution to the disk budget, however, amounts only to about $\sim 10\%$ of the
stellar mass-density. Adding up the present determination and the previously
 determined
stellar contribution yields the Galactic disk local density of star-like objects
$\rho_\odot\simeq 0.05\,\mvol$.

\bigskip

\begin{acknowledgements} The author is thankful to Xavier Delfosse, Adam Burgasser and Davy Kirkpatrick for helpful discussions about the observational data, and to the anonymous referee whom remarks helped clarifying the present paper.
I am also very grateful to the Astronomy Department and to the Miller Institute of the University of Berkeley, where this work was initiated, for a visiting professor position.
\end{acknowledgements}

\clearpage\eject

\bigskip

\begin{table}
\caption[]{L-dwarf and T-dwarf detections for different surveys.
In columns 4 and 5, "obs" is the observed number of objects while "MF1", "MF2"
and "MF4" denote the number of objects obtained with the IMF1, IMF2 and IMF4, respectively.
The references are: (1) Delfosse et al. (1999); (2) Basri et al. (2000); (3) Kirkpatrick et al. (1999);
(4) Kirkpatrick et al. (2000); (5) Burgasser (2000); (6) Tsvetanov et al. (2000);
(7) Leggett et al. (2000). The magnitudes are the ones in the standard UKIRT system. \label{table1}}
\bigskip
\begin{tabular}{lcccc}
\tableline
Survey & $\Omega$ & mag. lim. & L-dwarfs  & T-dwarfs )  \\
& (deg$^2$) &   & (J-K$\simgr 1.0$) & (J-K$\simle 1.0$)  \\
&  & & obs/MF1/MF2/MF4 & obs/MF1/MF2/MF4 \\
\hline \\
{\bf DENIS$^{(1),(2)}$}  & 240 & K$\le 13.5$  & 5/13/6/7 &   \\
             &     &  &       &   \\
{\bf 2MASSI$^{(3)}$} & 371 & K$\simle 15$; J-K$>$1.3 & 19/35/18/22 &   \\
             &     &  &       &   \\
{\bf 2MASSII$^{(4)}$}& $\sim$4000 & K$\simle 15$, J-K$>$1.3 & &   \\
             &     & d$\le 25$ &  42/50/27/30  &   \\
             &     & {\rm whole survey} & $\sim 100$/400/180/240 \\
{\bf 2MASSII$^{(5)}$}& 18360 & K$\simle 14.5$, J-H$\le$0.3, H-K$\le$0.3 &  &  13/88/27/42 \\
             &       &  &       &   \\
{\bf SDSS$^{(6)}$}   & 130   & z$^\star\simle 19$, J$\simle 16$  &  & 2/6/2/3 \\
             &       &  &       &   \\
{\bf SDSS$^{(7)}$}   & 225   & z$^\star\simle 19$, J$\simle 16$  &  &  3/10/3/5
\\
             &       &        &       &   \\
\tableline
\end{tabular}
\end{table}

\newpage

\vfill\eject

\bigskip

\begin{table}
\caption[]{Disk present-day brown dwarf density in the Galactic disk calculated
with the BDIMF IMF2 (upper row) and IMF4 (lower row). The scale height for the determination of the surface density is 250 pc. \label{table2}}
\bigskip
\begin{tabular}{lcccc}
\tableline
&  $n_\star$ & $\rho_\star$ & $\Sigma_\odot$   \\
& (pc$^{-3}$) & (M$_\odot$ pc$^{-3}$) & (M$_\odot$ pc$^{-2}$) \\
\hline \\
\mbox{}\hspace{0.2cm} BD systems & 0.09 & $3.0\times 10^{-3}$ & 1.5 \\
\mbox{}\hspace{0.2cm}            & 0.28 & $4.6\times 10^{-3}$ & 2.3 \\
\mbox{}\hspace{0.2cm} BD-BD companions & 0.03 & $1.0\times 10^{-3}$ & 0.5 \\
\mbox{}\hspace{0.2cm}            & 0.09 & $1.5\times 10^{-3}$ & 0.8 \\
\mbox{}\hspace{0.4cm} ($X_{BD}=0.3$, $q(m)=1$) & & & \\
\mbox{}\hspace{0.2cm} BD-star companions &  1.0-6.0 $\times 10^{-3}$ & $\simle 0.5\times 10^{-3}$ & $< 0.5$ \\
\mbox{}\hspace{0.4cm} ($f_{BD}=$0.05-0.3) & & & \\
\hline \\
\mbox{}\hspace{0.5cm} all BDs  & $\sim 0.12$  & $\sim (4.0$-4.5)$\times 10^{-3}$ &
$\sim 2$ \\
\mbox{}\hspace{0.5cm}      & $\sim 0.37$  & $\sim (6.0$-6.5)$\times 10^{-3}$ &
$\sim 3$ \\ 
\tableline
\end{tabular}
\end{table}
\clearpage\eject

\begin{figure}

\centerline {\bf FIGURE CAPTIONS}
\vskip1cm

\caption[]{Mass distribution for objects with mass 0.001$\le m/\msol\le 0.072$ obtained with the power law IMF1
(dash line), the power law IMF4 (dash-dot) and the lognormal IMF2 (solid line) for $J\le 22$ and $\Omega=$ 10 sq.deg.}
\label{chab2_fig_mass.ps}
\end{figure}

\begin{figure}
\caption[]{Age distribution as a function of J-K colors for the reference survey conditions, for IMF1 (upper figure) and IMF2 (lower figure), respectively}
\label{chab2_fig_age_col.ps}
\end{figure}

\begin{figure}
\caption[]{Brown dwarf distribution as function of J-K colors for IMF1 (dash line) and IMF2 (solid line) for $J\le 22$}
\label{chab2_fig_color.ps}
\end{figure}

\begin{figure}
\caption[]{Brown dwarf effective temperature distribution}
\label{chab2_fig_teff.ps}
\end{figure}

\begin{figure}
\caption[]{Predicted systemic brown dwarf discovery function obtained with the IMF1 (long-dash line), the IMF4 (dash-dot line) and the IMF2 (solid line). The dotted and short-dash lines display the T-dwarf contributions for IMF1 and IMF2, respectively. The error bars illustrate 3$\sigma$ Poisson error bars.}
\label{chab2_figIMF_col.ps}
\end{figure}

\begin{figure}
\caption[]{Predicted systemic brown dwarf discovery function obtained with the IMF1 and IMF2, for a constant SFR (eqn.(\ref{eqnsfr1}); long-dash and solid lines) and an exponentially-decreasing SFR
(eqn.(\ref{eqnsfr2}); dot and short-dash line)}
\label{chab2_figSFR_col.ps}
\end{figure}

\begin{figure}
\caption[]{Predicted brown dwarf discovery function obtained with IMF2 and a constant SFR for the unresolved systems (solid line) and for the resolved objects for $X_{BD}=0.30$, for a companion
uniform mass-distribution (eqn.(\ref{eqnpq1}); long-dash line),
gaussian mass-distribution (eqn.(\ref{eqnpq2}); dotted line) and equal-mass
distribution  ($q(m)=1$; short-dash line)}
\label{chab2_figBIN.ps}
\end{figure}

\vfill\eject

\vfill\eject
\begin{figure}
\begin{center}
\epsfxsize=190mm
\epsfysize=200mm
\epsfbox{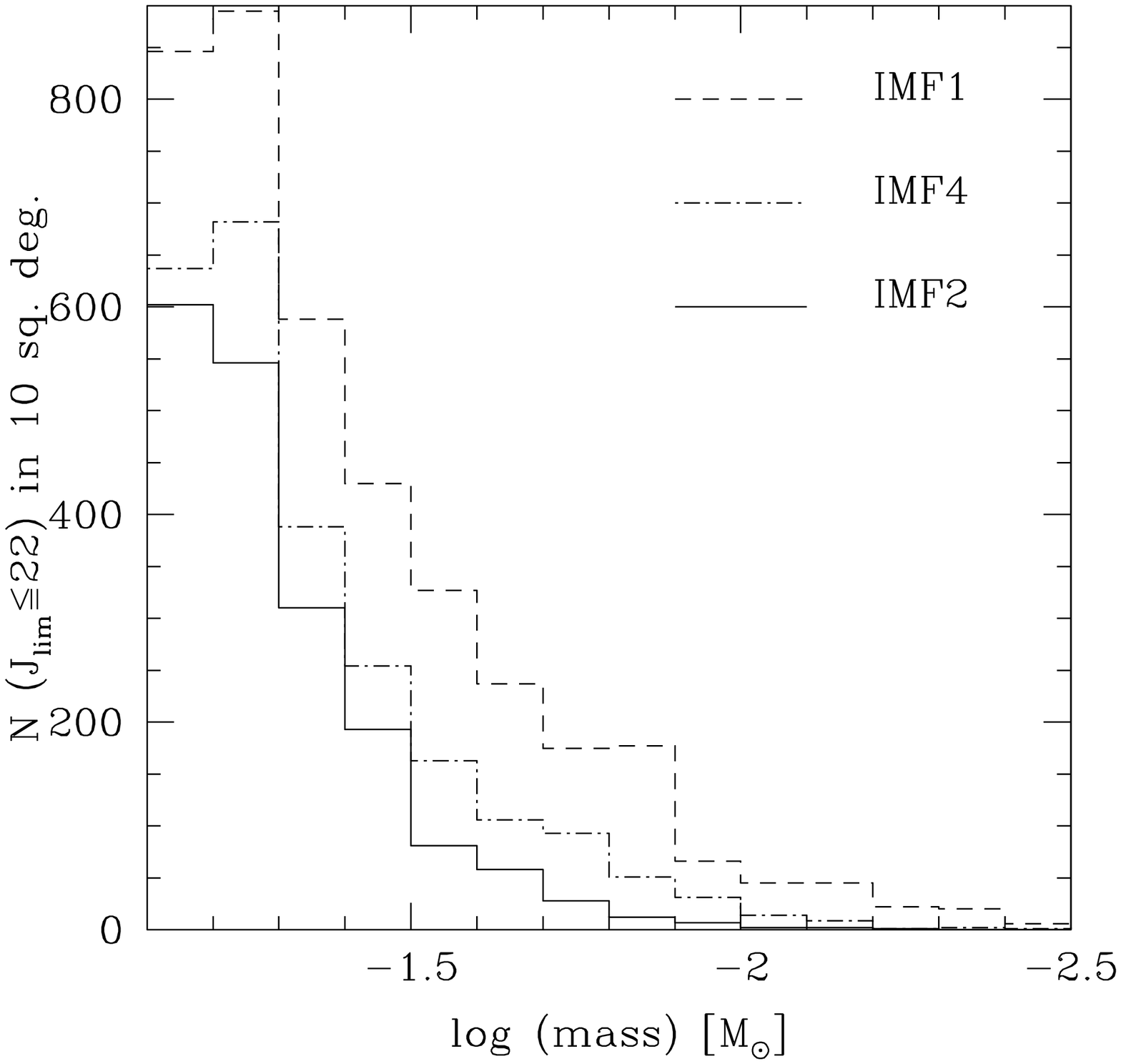}
\end{center}
\end{figure}

\vfill\eject

\vfill\eject
\begin{figure}
\begin{center}
\epsfxsize=190mm
\epsfysize=200mm
\epsfbox{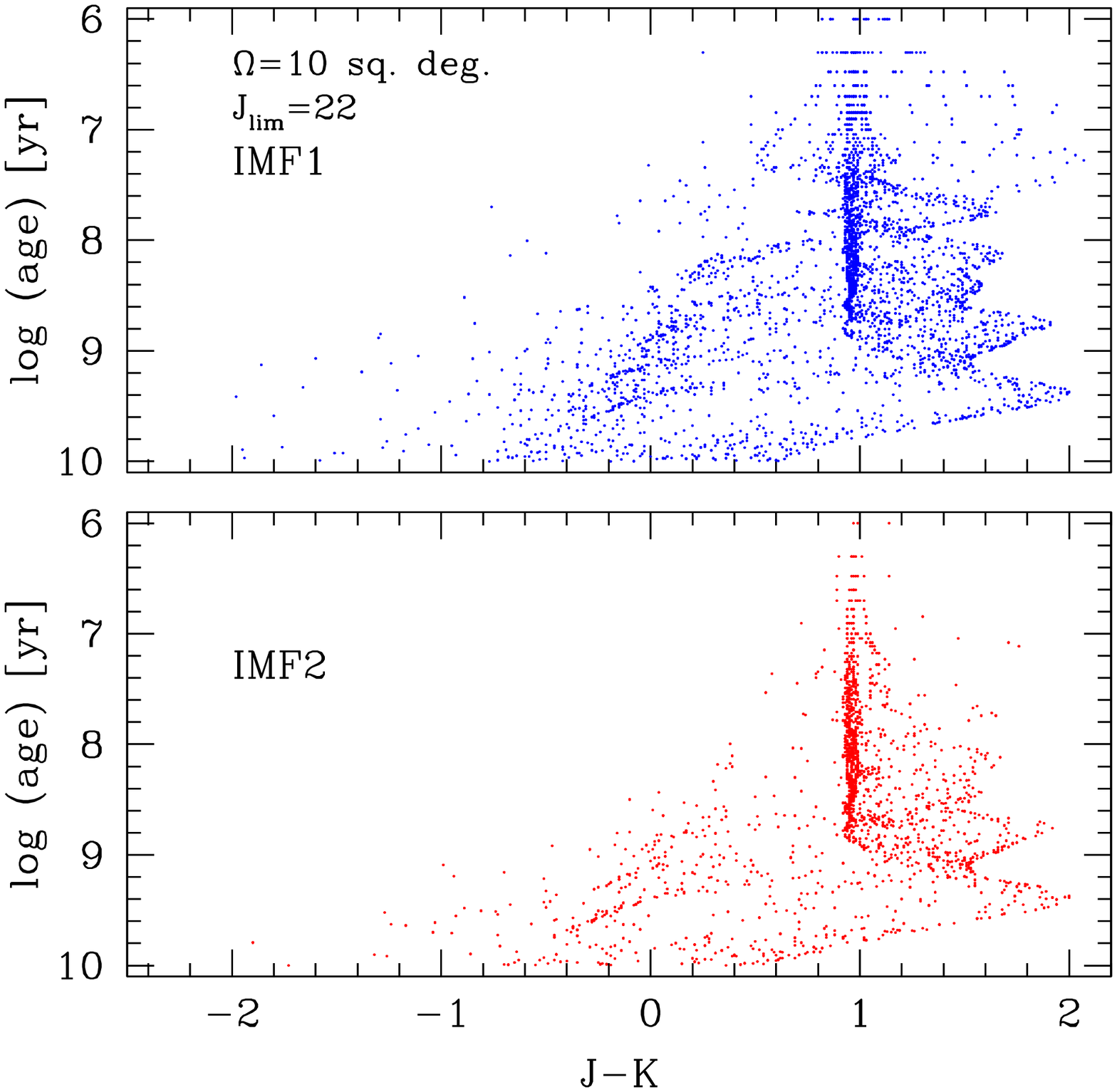}
\end{center}
\end{figure}

\vfill\eject

\vfill\eject
\begin{figure}
\begin{center}
\epsfxsize=190mm
\epsfysize=200mm
\epsfbox{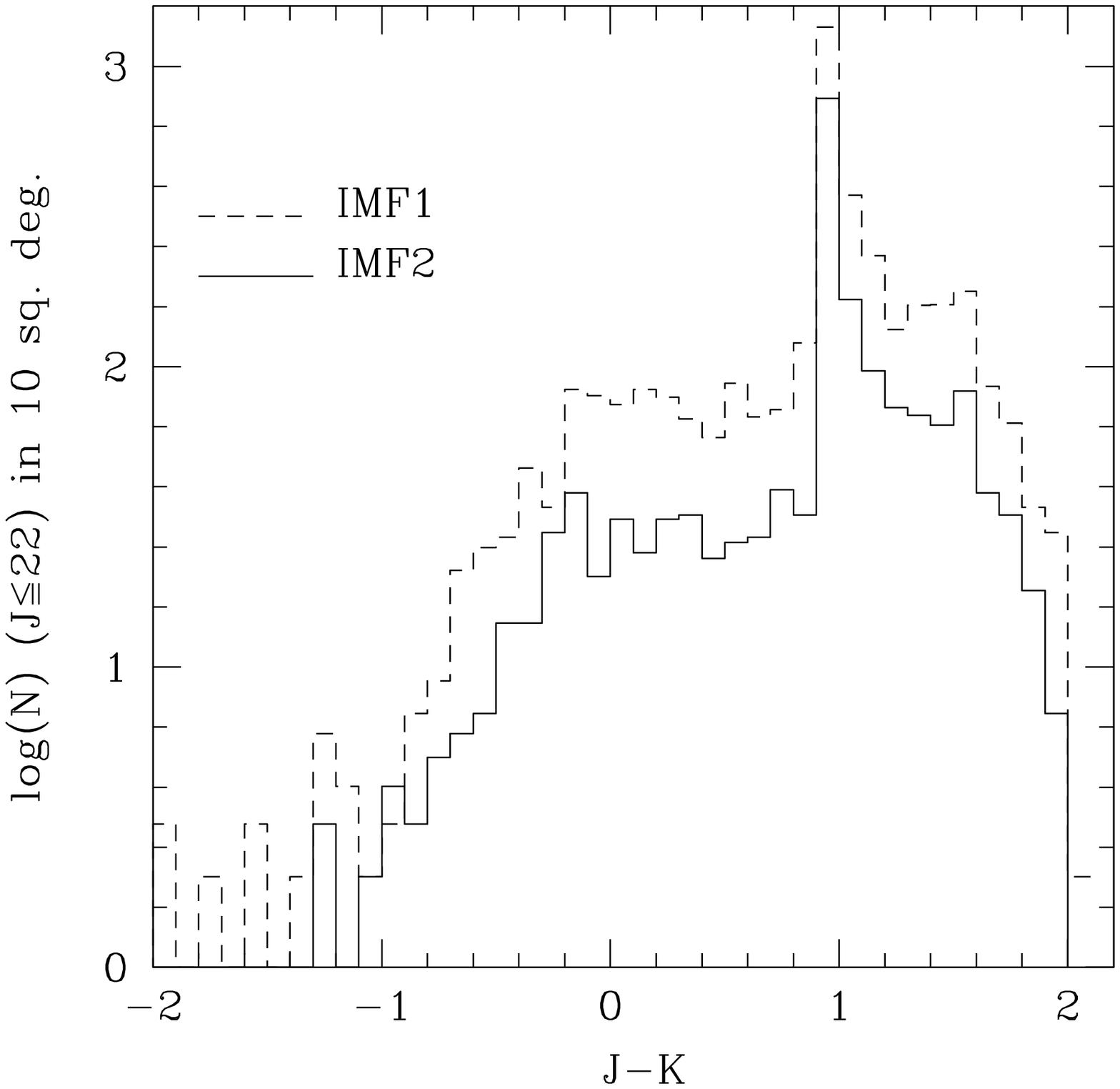}
\end{center}
\end{figure}
\vfill\eject

\vfill\eject
\begin{figure}
\begin{center}
\epsfxsize=190mm
\epsfysize=200mm
\epsfbox{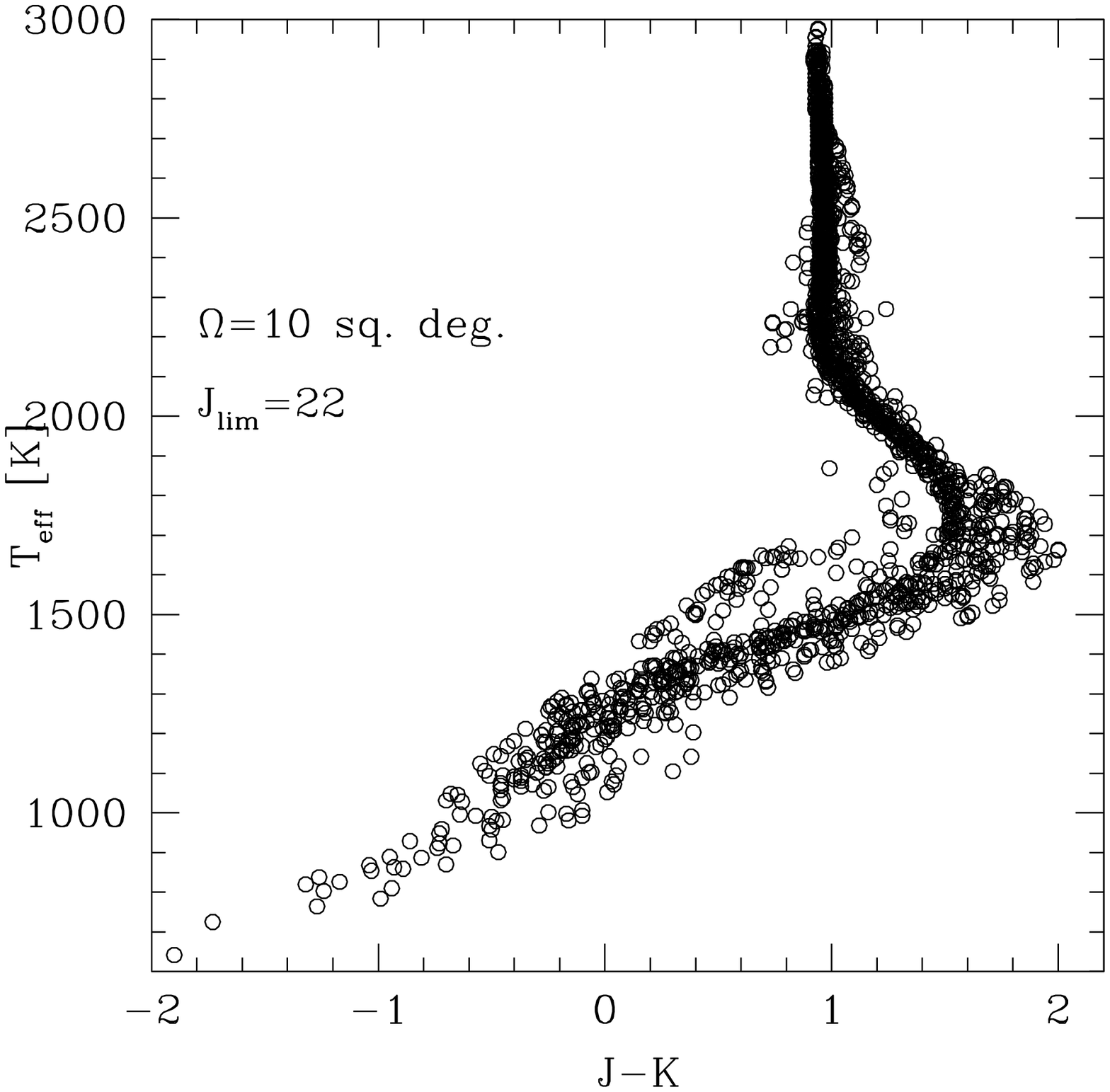}
\end{center}
\end{figure}

\vfill\eject

\vfill\eject
\begin{figure}
\begin{center}
\epsfxsize=190mm
\epsfysize=200mm
\epsfbox{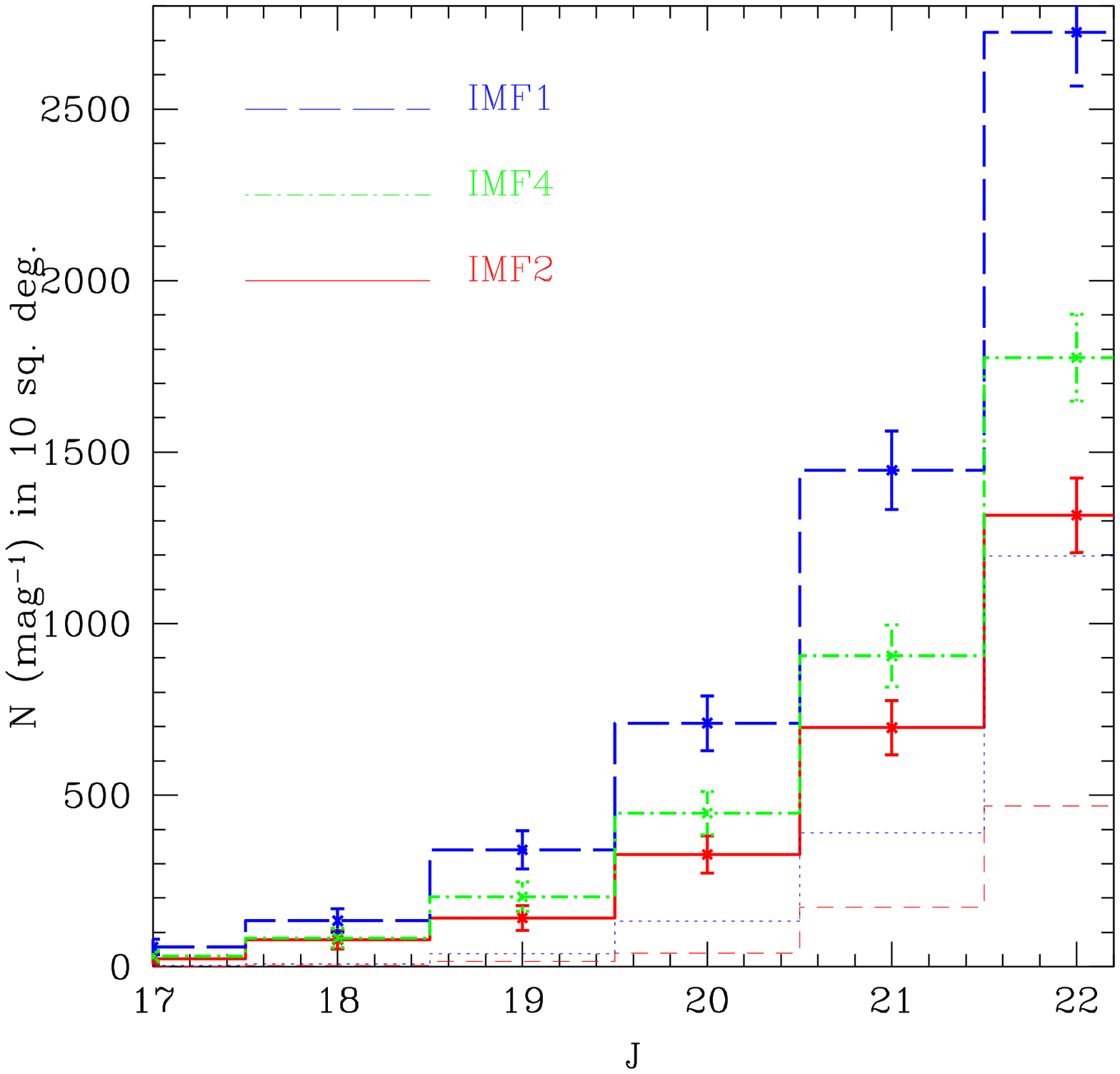}
\end{center}
\end{figure}

\vfill\eject

\begin{figure}
\begin{center}
\epsfxsize=190mm
\epsfysize=200mm
\epsfbox{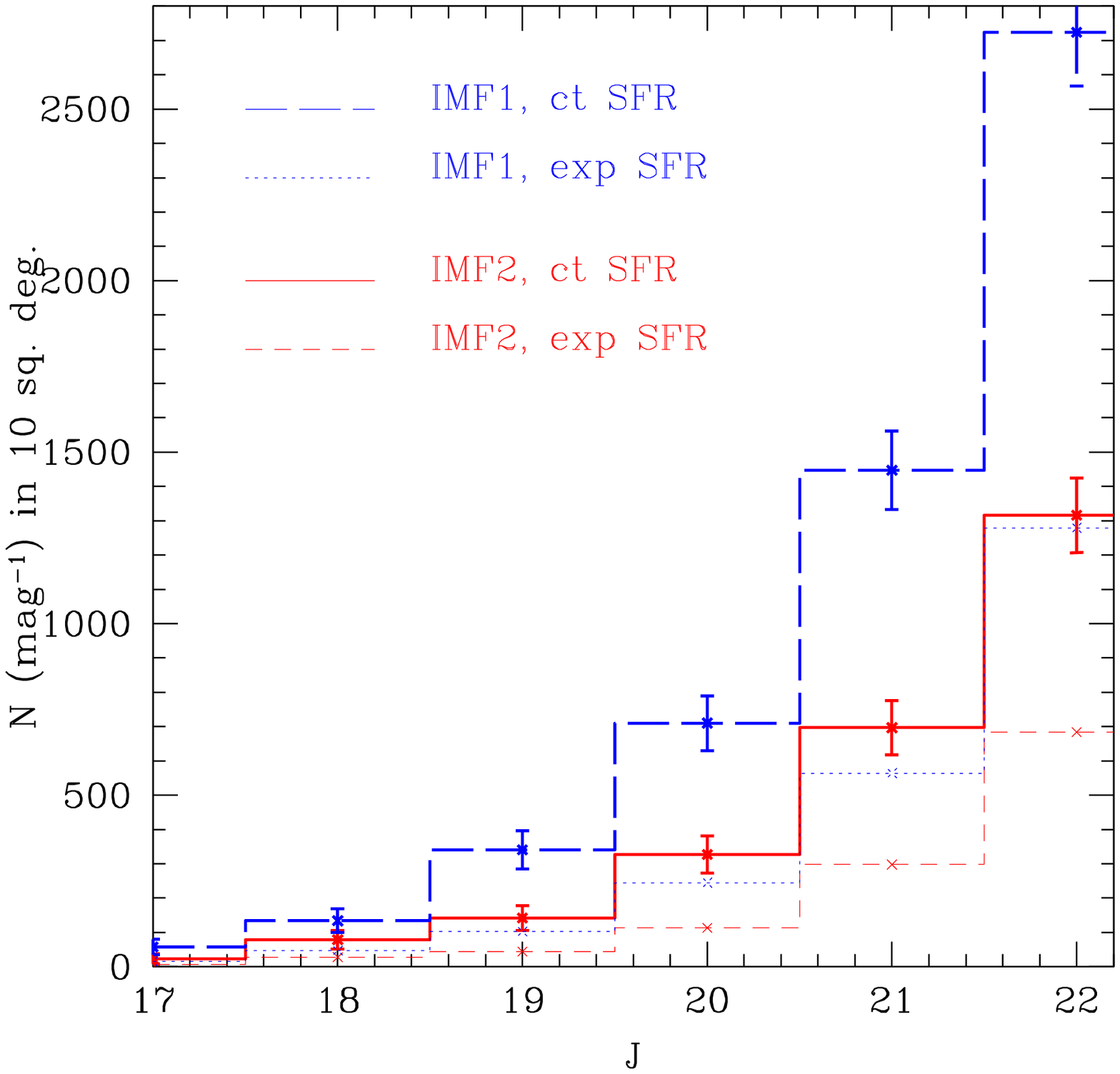}
\end{center}
\end{figure}

\vfill\eject

\begin{figure}
\begin{center}
\epsfxsize=190mm
\epsfysize=200mm
\epsfbox{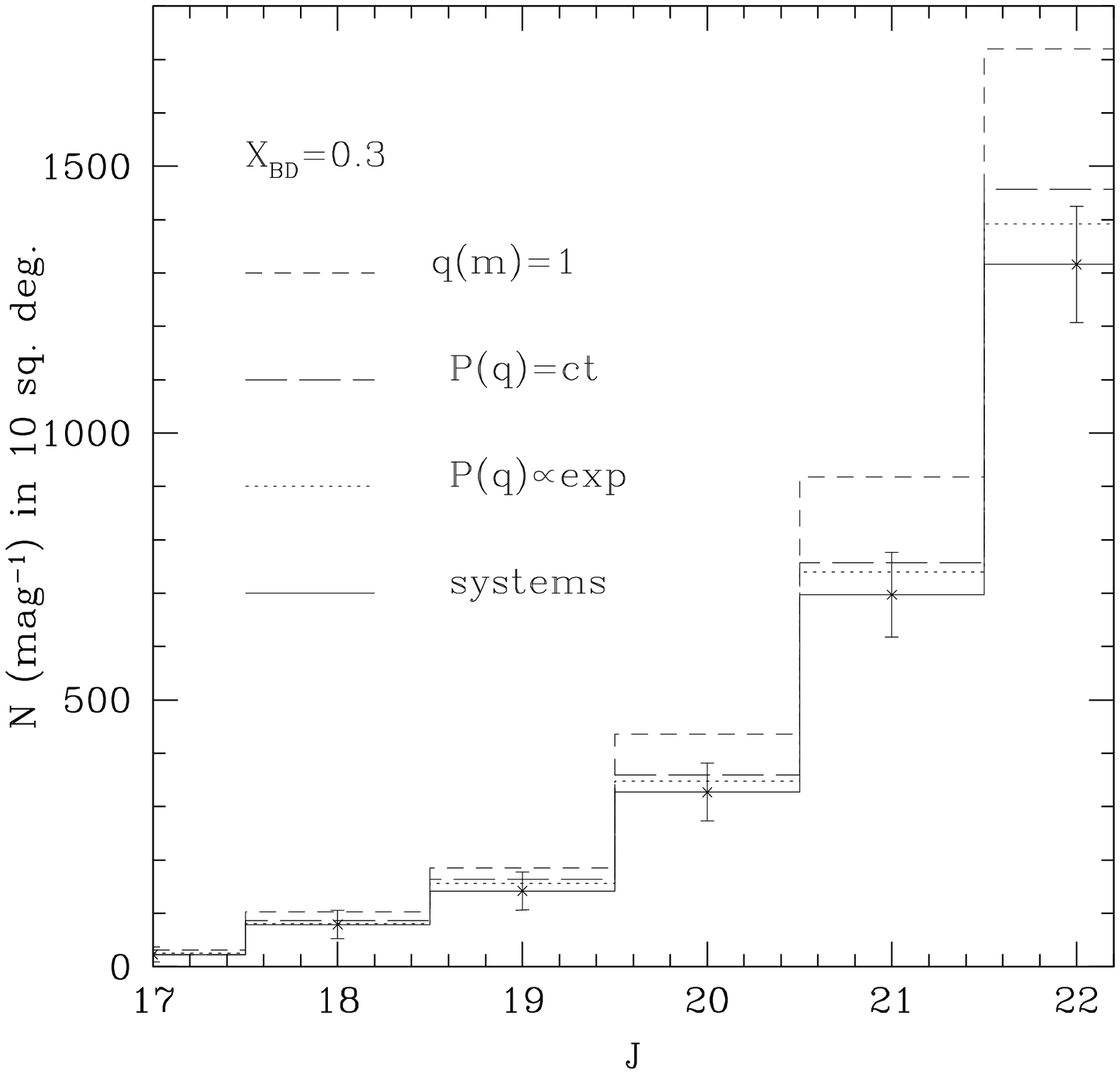}
\end{center}
\end{figure}

\vfill\eject

\end{document}